# Physics-embedded neural computational electron microscopy for quantitative 4D nanometrology

Hao-Jin Wang[1#], Liqun Shen[1#], Xin-Ning Tian[1#], Lei Lei[2#], Kexin Wang[1], Grigore Moldovan[3], Marc-Georg Willinger[4], Zhu-Jun Wang[1*]

1 School of Physical Science and Technology & Shanghai Key Laboratory of High-resolution Electron Microscopy, ShanghaiTech University, Shanghai, China.

2 Center for Transformative Science, ShanghaiTech University, Shanghai, China.

3 Point Electronic Gmbh, Halle (Saale), Saxony-Anhalt, Germany

4 Department of Chemistry, Technical University of Munich, Garching, Germany.

Corresponding Author: wangzhj3@shanghaitech.edu.cn

**Abstract**

The fusion of rigorous physical laws with flexible data-driven learning represents a new frontier in scientific simulation, yet bridging the gap between physical interpretability and computational efficiency remains a grand challenge. In electron microscopy, this divide limits the ability to quantify three-dimensional topography from two-dimensional projections, fundamentally constraining our understanding of nanoscale structure-function relationships. Here, we present a physics-embedded neural computational microscopy framework that achieves metrological three-dimensional reconstruction by deeply coupling a differentiable electron-optical forward model with deep learning. By introducing a 'Vision Field Transformer' as a high-speed, differentiable surrogate for physical process analysis simulations, we establish an end-to-end, self-supervised optimization loop that enforces strict physical consistency with hardware geometry. This synergy enables single-shot, quantitative three-dimensional nanometrology with precision comparable to atomic force microscopy but at orders of magnitude higher throughput. Furthermore, we demonstrate the capability for 'four-dimensional (three-dimensional real space + time)' in-situ characterization by tracking the dynamic evolution of surface nanostructure during copper-redox, revealing hidden crystallographic kinetics invisible to conventional imaging. Our work not only redefines the limits of scanning electron microscopy but also establishes a generalizable archetype for solving ill-posed inverse problems across physical sciences, unlocking the full potential of simulation as a third pillar of discovery.

**Introduction**

Physical simulation has become the third major pillar of science and engineering,

alongside theory and experiment[1, 2]. The field is now undergoing a clear shift in paradigm: on one hand, traditional methods grounded in physical laws are rigorous and testable, but often come with a high computational cost; on the other hand, data-driven methods based on neural networks are efficient and can fit complex patterns well, yet they often lack physical constraints and therefore offer limited interpretability and robustness when conditions change[3]. A central question, then, is how to combine physical trustworthiness with data efficiency—and build new simulation/inversion frameworks that are both fast and highly reliable.

Three-dimensional (3D) surface imaging in electron optics is governed by complete classical physics, yet it faces severe computational bottlenecks due to the complexity of electron scattering and trajectory evolution[4, 5]. This makes it an ideal testbed for the "physics–data fusion" paradigm. At the micro- and nanoscale, accurately revealing the 3D structure of materials and biological systems is essential for understanding how function emerges[6, 7, 8]. However, as microscopy moves from "capturing high-quality two-dimensional (2D) images" toward "quantitative 3D topography measurement"[9], the missing depth information in the imaging process becomes a key barrier: existing SEM images are still, in essence, intensity projections of 3D space onto a 2D plane, and they do not directly provide a height field with metrological meaning. To address this challenge, we introduce a closed-loop 3D metrology framework that tightly couples a differentiable electron-optics forward model with deep-learning-based inversion, which we name ECLIPS (Electron-contrast–constrained Learning with an Interpretable Physics Surrogate). ECLIPS aims to deliver high-fidelity reconstruction of microscopic 3D topography without compromising imaging efficiency. By design, reconstruction quality should be bounded primarily by hardware performance rather than algorithmic or procedural confounders, thereby providing a high-precision, high-throughput, and verifiable pathway for quantitative microscopy metrology.

As a cornerstone tool for micro/nanoscale characterization, scanning electron microscopy (SEM) is widely used for both static structure inspection and tracking dynamic processes, thanks to its intuitive imaging and convenient operation[10]. SEM forms images by scanning a focused electron beam point by point and collecting secondary-electron (SE) and backscattered-electron (BSE) signals[11]. Although SE is sensitive to surface relief and BSE is more sensitive to material composition[12, 13], and although the relative geometry between the sample and detector can introduce directional shading and a sense of "3D appearance" to some extent[12], the output is still a 2D contrast image[14] without direct depth calibration. As a result, recovering accurate 3D surface topography from 2D SEM/BSE contrast is fundamentally limited by information loss caused by dimensionality collapse, leading to an ill-posed inverse problem[15, 16].

A range of existing 3D reconstruction approaches have attempted to tackle this inverse problem, yet it remains difficult to meet both high accuracy and high efficiency at the same time. Multi-view reconstruction acquires images from different viewing angles by tilting the specimen stage[17]; it is conceptually straightforward, but is limited by mechanical precision, operational space, and time cost, making fast and stable continuous measurement challenging[18, 19]. Methods inspired by optical photometric stereo attempt to infer geometry from "illumination changes"[20, 21], but electron motion and scattering in electromagnetic fields are far more complex than optical light paths[22, 23]. Directly transferring optical assumptions can introduce systematic physical bias and undermine metrological accuracy. Purely data-driven black-box learning can fit data, but without physical constraints it is often sensitive to changes in imaging conditions (such as voltage, working distance, detector relative position, material differences, and so on)[24], and thus struggles to deliver the stability and verifiability required for scientific metrology[3].

To bridge the gap between physical fidelity and reliable imaging, our goal is to build a general, robust, and interpretable 3D metrology reconstruction framework. ECLIPS makes a key choice at the signal level: instead of using SE, which are more easily affected by local electric fields and charging, we prioritize BSE, which can be efficiently collected even under weak bias and are better suited for building testable physical models[25, 26]. This strategy reduces uncontrolled field-induced effects on electron trajectories, providing a "cleaner" physical basis for constructing a high-confidence electron-optics forward model and a unified experimental–simulation reference frame.

Building on this, ECLIPS adopts a co-design of hardware and algorithms to enable high-accuracy, high-efficiency reconstruction within a single scan. On the hardware side, we integrate and carefully calibrate a four-quadrant (4Q) BSE detector inside the microscope, so that multiple BSE contrast images carrying directional information can be captured simultaneously in one scan, providing well-defined geometric boundary conditions for electron-optics modeling. On the algorithm side, ECLIPS uses a deep network to estimate the surface normal field (normal map) from multi-directional BSE contrast, and then employs a differentiable electron-optics forward model to render the normal map into simulated BSE contrast that can be directly compared with measurements. This forms a self-supervised analysis-by-synthesis loop: by minimizing the photometric consistency residual between simulation and observation, backpropagation drives the network to iteratively converge toward a physically consistent geometric solution[27].

After obtaining a high-quality normal field, ECLIPS further needs to convert it into a metrologically meaningful 3D depth field (depth map). Classical normal integration methods (such as solving the Poisson equation or path integration) can be highly sensitive—especially under noisy normals—to boundary conditions, deviations from integrability, and low-frequency drift, often leading to global shape drift or stripe-like artifacts[28, 29, 30]. We therefore introduce a physics-informed neural network (PINN) as the integrator[31]. By explicitly enforcing global geometric consistency, integrability, and necessary boundary constraints in the loss function, we achieve a robust mapping from normal map to depth map, significantly suppressing noise-induced low-frequency drift while preserving key geometric details.

The value of ECLIPS goes beyond static 3D metrology. Thanks to the efficiency of capturing multi-directional information in a single scan, and the consistency and verifiability enabled by the physics-constrained closed loop, the framework naturally extends to frame-by-frame 3D reconstruction of in situ dynamic processes, offering a feasible path toward "4D (3D + time)" metrology. For example, in the redox process on copper surfaces[32], ECLIPS can stably recover time-resolved 3D evolution and quantitatively resolve nanoscale dynamic structures such as step-flow driven by screw dislocations[33], revealing 3D geometric change mechanisms that are hard to capture reliably with traditional 2D contrast imaging. More broadly, the closed-loop design embodied by ECLIPS—"a differentiable physics forward model + neural-network inversion"—provides a transferable modeling paradigm for tackling ill-posed inverse problems across many physical systems: it retains computational efficiency while turning physical priors into differentiable, testable constraints, thereby improving the scientific credibility and metrological meaning of reconstruction results[31, 34, 35].

## Physics-embedded neural computational microscopy for quantitative three-dimensional nanometrology

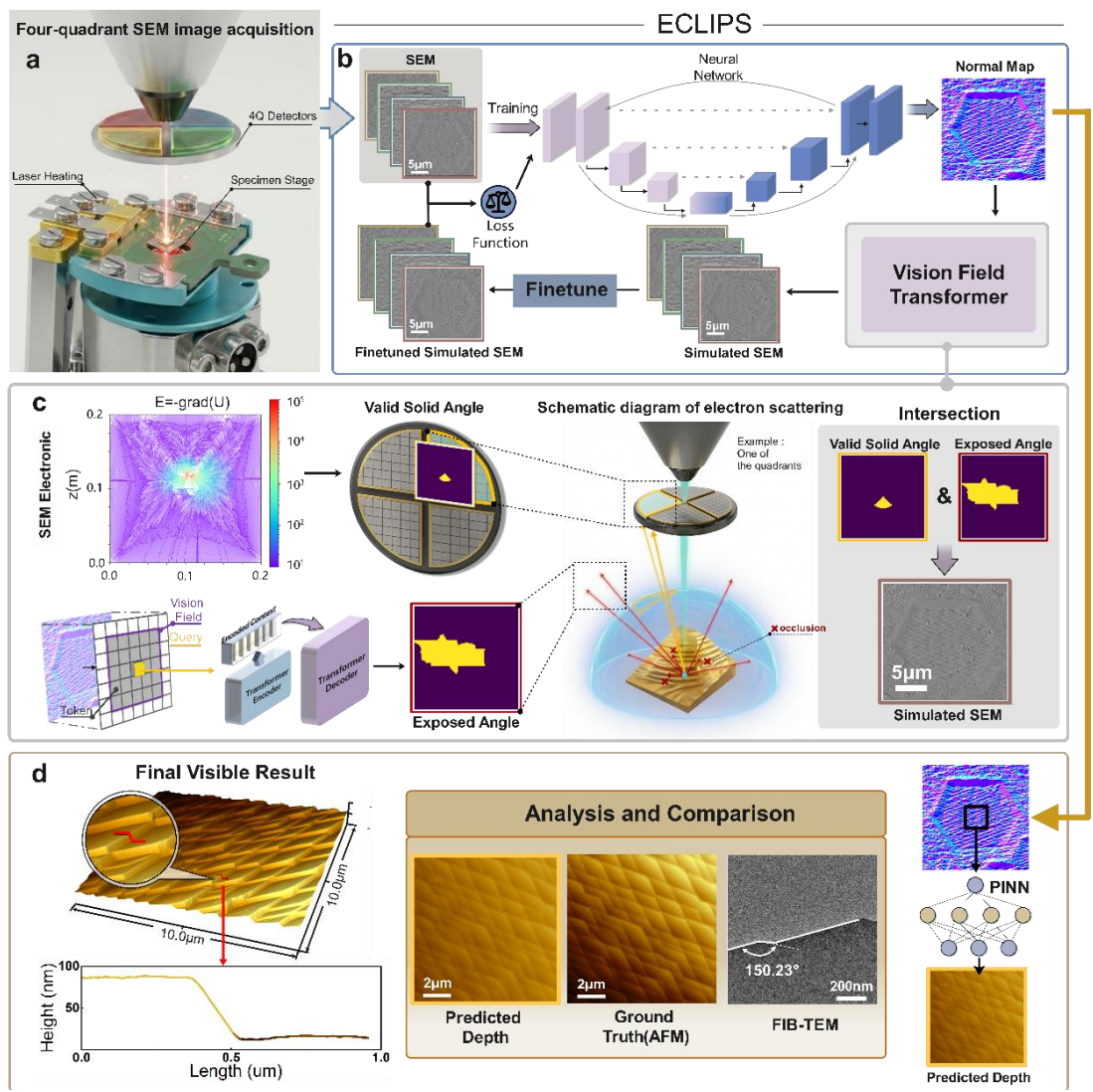

**Fig. 1 | ECLIPS: physics-embedded neural computational microscopy for quantitative three-dimensional SEM reconstruction.**

**(a)** Data acquisition. A customized 4Q-BSE detector mounted beneath the SEM pole piece records four azimuth-resolved BSE images in a single scan from the specimen on a geometry-calibrated stage (optionally with in situ actuation/heating). These four images form the multi-view SEM input stack for reconstruction. **(b)**, From BSE images to a normal map, then back to SEM for self-consistency. The 4Q-SEM stack from a is fed into a U-Net–style neural backbone to predict a surface normal map (right). The predicted normal map is then passed to the Vision Field Transformer (VFT) forward renderer to generate simulated SEM images corresponding to the same 4Q views. A finetuning module further adjusts the simulated SEM to match measurement appearance. The simulated/finetuned SEM is compared with the real SEM input to compute an image-consistency loss, and this loss is used to update the reconstruction network (closing the self-supervised loop indicated by the arrows). **(c)**, How the forward renderer synthesizes SEM contrast. Given the predicted surface geometry (from the normal map), the forward model determines, for each pixel and each quadrant, the valid solid angle set by the detector geometry and instrument settings, the exposed angle set by local surface visibility/occlusion, and their intersection (visible subset). This visible subset governs the quadrant-wise BSE intensity used to produce the simulated SEM images that feed back into the loss in **(b)**. **(d)**, From normal map to quantitative depth and validation. The optimized normal map from b is converted into a depth/height map using a physics-informed integrator (PINN block). The resulting 3D topography is reported as a rendered surface and line profiles, and is validated by cross-modal comparison to AFM height maps and FIB–TEM cross-sectional measurements (analysis panel). Scale bars as indicated.

To overcome the inherent limits of conventional surface-topography methods in accuracy, speed, and applicability to non-equilibrium dynamics, we developed the ECLIPS framework. ECLIPS tightly couples a hardware imaging system customized for backscattered-electron (BSE) imaging with a differentiable electron-optics model, aiming for an end-to-end quantitative inversion from 2D BSE contrast images to 3D surface topography. As shown in **Fig. 1**, ECLIPS establishes a complete closed loop from hardware acquisition to 3D reconstruction, integrating electron-optics simulation with self-supervised optimization. The workflow consists of three coordinated components: **Fig. 1a** shows the hardware setup and experimental data acquisition; **Fig. 1b** presents the core deep-learning module; and **Fig. 1c** illustrates the physics simulation architecture based on electron-optics principles. By integrating these modules in a self-consistent way, we achieve accurate reconstruction from multi-directional BSE contrast images to a 3D topography matrix, and we further confirm metrology-level accuracy and strong performance at the nanoscale through multimodal cross-validation with AFM and FIB-TEM (**Fig. 1d**).

First, on the acquisition side, we designed and integrated a 4Q-BSE detector based on electron-optics imaging geometry, enabling the system to capture, in a single scan, four BSE images that carry different azimuthal-direction information (**Fig. 1a**). The detector is placed beneath the SEM pole piece and forms a fixed relative geometry with the specimen stage. By calibrating the detector's spatial position and key operating parameters (such as accelerating voltage and working distance), we obtain high signal-to-noise BSE measurements while also establishing accurate geometric boundary conditions for later modeling. This aligns experimental observations and electron-optics forward simulation within a shared reference frame. Notably, to support external fields and environmental disturbances that commonly arise in in situ studies, the specimen-stage module further integrates a local laser-heating unit and provides reserved/compatible interfaces for controlled-atmosphere (environmental) imaging and control (**Fig. 1a**). This allows the 4Q acquisition scheme to extend to 3D topography tracking during dynamic processes under controlled gas and heating conditions. Together, the geometric calibration and system integration provide well-defined physical boundary conditions for subsequent BSE physics simulation, electron-trajectory prediction, and quantitative 3D reconstruction—forming the foundation for metrological inversion.

Building on the hardware and geometric calibration, we constructed a matching computational inversion module that converts the multi-directional information from the 4Q-BSE signals into geometric representations suitable for 3D reconstruction. As shown in **Fig. 1b**, in the core deep-learning module we use U-Net-Style network as the backbone network[36, 37, 38], learning an initial mapping from multi-azimuth BSE contrast images to a surface normal map. Because inferring 3D geometry from 2D intensity observations is inherently ill-posed[15], a purely data-driven model can easily absorb empirical biases that strongly depend on imaging conditions. To obtain a stable initialization and speed up convergence, we first adopt a supervised pretraining strategy: we compute normal fields from AFM height maps of the same region as training labels, and use machine-learning-based registration to achieve strict spatiotemporal alignment[39, 40]. However, AFM data inevitably contains error sources such as tip-trailing artifacts and registration residuals[41], meaning the supervisory signal itself is noisy. In addition, purely data-driven mappings often generalize poorly across changes in operating conditions such as accelerating voltage, making pretrained results relatively coarse and limiting robustness across experiments.

To address this domain gap problem, we introduce an electron-optics forward model into ECLIPS and build a physics-guided deep-learning simulation pipeline (**Fig. 1c**). Based on the precise spatial calibration described above and electron-optics principles, we build the BSE Simulation Model (BSESM), which can rigorously simulate BSE emission and flight trajectories at the pixel level and provide a forward mapping from 3D surface topography to 2D BSE contrast distributions. However, conventional BSESM computation is extremely expensive: under typical computing resources, a single iteration can take tens of hours, and the forward process is non-differentiable, preventing seamless integration into gradient-based training.

To overcome this bottleneck, we propose a Vision Field assumption based on surface statistical properties. Through a rigorous probabilistic derivation, this assumption shows that if electrons are not occluded within a specific local radius r centered at the emission point (the Vision Field), then the conditional probability of encountering occlusion in the far field quickly decays to a negligible threshold $\delta_{out}$. This result mathematically reduces a globally dependent 3D ray-tracing problem into a lower-dimensional 2D feature-query problem that depends only on local neighborhood information. Guided by this, we design the Vision Field Transformer (VFT) as an efficient surrogate for BSESM, using attention mechanisms to efficiently resolve local 3D occlusion and visibility on a 2D plane[42, 43]. Although the statistical assumption has boundary limitations in theory at low emission angles (low direction thresholds), analysis combining detector geometry and field-driven trajectories shows that such low-angle electrons cannot enter the effective solid angle of the 4Q detector in the first place. Therefore, while maintaining strict physical fidelity for the effective imaging signal, VFT compresses the per-simulation time from ~20 hours to seconds and makes the full pipeline differentiable, enabling the physics forward operator to be embedded into deep-network training.

We then embed VFT as a differentiable Forward Model into the U-Net reconstruction framework to form a physics-constrained self-supervised loop: U-Net takes multi-azimuth BSE images as input and predicts a normal field; the normal field is rendered by the Forward Model into simulated BSE contrast; and a photometric-consistency loss is computed against the measured images. During training, the error signal backpropagates to update U-Net parameters, forcing the predicted normal field to progressively approach the true geometry under physical constraints. Importantly, even with detailed electron-optics modeling and the Vision Field approximation, small systematic differences between simulation and real measurements can still remain due to complex surface scattering behavior (a sim-to-real gap)[44, 45]. To further reduce this residual, we add a lightweight neural refinement module after the Forward Model. Using joint features from the normal map and the initial simulated image, it adaptively predicts and corrects nonlinear distortions between simulation and measurement. This acts as a data-driven compensation layer for physical-model error, further improving image consistency. In addition, for use cases that are extremely latency-sensitive—such as large-field, high-throughput statistics or online monitoring—we build a purely deep-learning fast simulator (Deeplearning Simulation Model, DSM) on top of the Forward Model[46], distilling physical knowledge from BSESM/VFT to further reduce the computational complexity of the reconstruction framework.

To rigorously validate the reliability of ECLIPS, we use atomic force microscopy (AFM) and focused-ion-beam transmission electron microscopy (FIB-TEM) as ground truth direct measurements and conduct multimodal cross-checking (**Fig. 1d**). Specifically, we need to solve for a 3D depth field (Depth Map) from the optimized normal map. Unlike traditional Poisson-based solvers or path integration methods, in the micro/nanoscale regime studied here (overall flat samples with small local height variations), classical methods show clear limitations: Poisson-based integration is highly sensitive to boundary conditions and noise[28, 29], and small errors can accumulate into global shape drift or tilt distortion; path integration requires a strictly conservative field, and directly integrating noisy predicted normals often produces severe stripe-like artifacts[30]. We therefore propose and adopt a physics-informed neural network (PINN) integration architecture (Physics-Informed Neural Network, PINN)[31], enabling a robust mapping from Normal Map to Depth Map. In the loss function, we explicitly include PDE-residual constraints. This allows deep learning to capture the (potentially nonlinear) relationship between the normal field and the height field while preserving both global geometric consistency and local detail accuracy. Finally, through pixel-level registration and multimodal comparison with AFM and FIB-TEM results (see **Fig. 1d** Analysis and Comparison), we verify the high accuracy and low-noise properties of the AI reconstruction (We will discuss in following), obtaining a 3D surface-topography matrix that is highly consistent with experimental observations and theoretical expectations (as shown in **Fig. 1d** Final Result and the line profiles).

**Electronic-optical physical modeling and high-confidence neural surrogate validation**

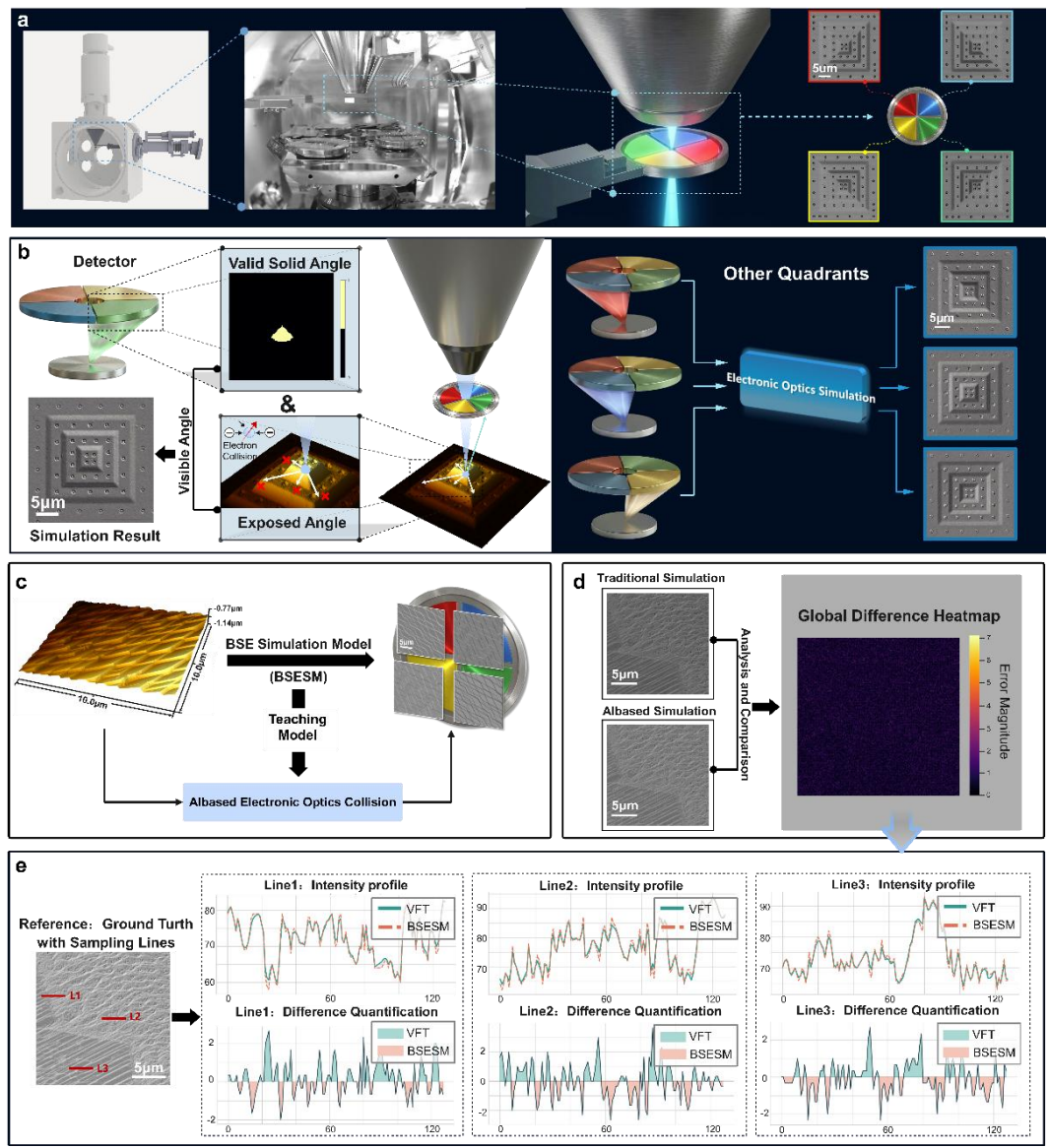

**Fig. 2 | Electronic-optical simulation framework and neural surrogate acceleration.**

**(a)**, Geometry-calibrated hardware baseline for multi-azimuth BSE acquisition. The SEM column illuminates the specimen while a customized 4Q detector, registered to the specimen stage, captures four directional BSE images in a single scan; the colored quadrant indicator and the four example images illustrate the one-to-four mapping from the physical detector quadrants to the measured 4Q-BSE set. **(b)**, Electron-optical forward simulation under the calibrated geometry (shown as a "one-quadrant example" on the left and "other quadrants analogously" on the right). Left: the detector geometry defines a Valid Solid Angle (yellow region), i.e., emission directions that would reach this detector quadrant without considering occlusion. The sample topography further defines an Exposed Angle, i.e., emission directions that are not blocked by surrounding relief; red crosses mark representative emission directions that are occluded (blocked by local topography) and therefore cannot contribute to the detector signal. The Visible Angle is the intersection of Valid Solid Angle and Exposed

Angle (indicated by "&" and the vertical "Visible Angle" bracket), which then determines the simulated intensity for this quadrant (arrow to the "Simulation Result"). Right: the same procedure is applied to the other three quadrants (three colored cones), all feeding into the same electronic-optics simulation module to produce the corresponding three simulated quadrant images (arrows from "Other Quadrants" → "Electronic Optics Simulation" → output images). **(c)**, Neural surrogate learning for accelerating the forward model. A 3D topography input is rendered by the high-fidelity BSE simulator (BSESM, used as a teaching model), and the resulting paired data supervise an AI-based electronic-optics collision surrogate, forming the teacher→student (BSESM→AI) distillation pathway (arrows). **(d)**, Fidelity check of the surrogate renderer. The traditional/BSESM simulation and the AI-based simulation are compared side-by-side; their pixel-wise discrepancy is summarized as a global difference heatmap (arrow "Analysis and Comparison" → heatmap). **(e)**, Line-based quantitative evaluation. Sampling lines (L1–L3) are drawn on a reference image (left), then mapped to paired intensity profiles (VFT/AI vs BSESM) and corresponding difference quantification for each line (arrows from the reference with lines → profile panels).

Building on the ECLIPS hardware architecture described above, we next focus on the electron-optics simulation and the AI-accelerated validation system constructed on this geometric reference (**Fig. 2**). Using the precisely calibrated spatial relationship between the detector and the specimen shown in **Fig. 2a**, we first establish strict alignment between physical simulation and experimental acquisition within the same coordinate frame. This geometric consistency ensures that the multi-azimuth BSE signals captured by the detector not only exhibit strong directional shadowing effects, but also provide reliable experimental ground truth for validating the physical fidelity of electron-optics simulation.

On top of this hardware reference, we develop a first-principles BSE imaging simulator, termed the BSE Simulation Model (BSESM), as illustrated in **Fig. 2b**. To disentangle the imaging mechanism in a complex electromagnetic environment, we decouple the simulation into two key stages: the Valid Solid Angle and the Exposed Angle. The former depends only on macroscopic microscope parameters (such as the electric-field distribution, working distance, and accelerating voltage of the incident electrons). It describes the range of emission directions that can be received by the detector in the absence of occlusion, and can be precomputed as a lookup table. Building on that, the Exposed Angle is determined by the local 3D geometry of the specimen. It characterizes a visibility screening of these "receivable" emission directions under occlusion by the actual surface topography, reflecting the effective emission paths after being blocked by surrounding features. As a result, the final Visible Angle can be understood as the remaining visible subset of the Valid Solid Angle after applying local occlusion constraints. Its magnitude and directional distribution then determine the BSE signal intensity at each pixel for different detector quadrants.

Although BSESM can generate highly accurate physical ground truth, it relies on ray tracing and Monte Carlo integration, which have extremely high computational complexity and are difficult to parallelize. For example, generating a single 1024 × 1024 image takes more than 20 hours on a high-performance workstation (e.g., an AMD 9950X CPU). More critically, this discrete sampling process is inherently non-differentiable and cannot be directly embedded into gradient backpropagation for deep neural networks, creating a major barrier to end-to-end optimization.

To overcome this computational bottleneck, we propose a neural surrogate model based on the Vision Field assumption—the Vision Field Transformer (VFT)—and we rigorously validate its fidelity (**Fig. 2d**). VFT aims to approximate the physical mapping of BSESM with high accuracy in the form of a deep neural network. To train this model, we construct a large database of synthetic 3D topographies and introduce multi-scale geometric perturbations and surface-roughness augmentation to cover a broad range of material properties and imaging conditions. For optimization, we use the Muon optimizer together with a hard example mining strategy[47, 48], which improves convergence in a high-dimensional parameter space and substantially strengthens robustness in complex edges and low signal-to-noise regions. **Fig. 2d–e** present quantitative comparisons between VFT and conventional BSESM: for the same 3D topography input, the simulated image generated by VFT (**Fig. 2d, bottom left**) is visually almost indistinguishable from the BSESM ground truth (**Fig. 2d, top left**). Pixel-wise difference heatmaps (**Fig. 2d, right**) and line-profile intensity analysis (**Fig. 2e**) further show that the relative error is very small and is mainly confined to high-frequency edge regions, confirming the high fidelity of the AI-based simulation.

More importantly, by leveraging the parallelism of convolutional computation, VFT preserves physical accuracy and gradient continuity while reducing computational complexity from $o(n^4)$ to $o(n^2)$. This successfully compresses hours of physical simulation into millisecond-level inference, making VFT the core engine that enables the full closed-loop reconstruction system.

**First-principles electron-optical modeling and physics-data coupled reconstruction framework**

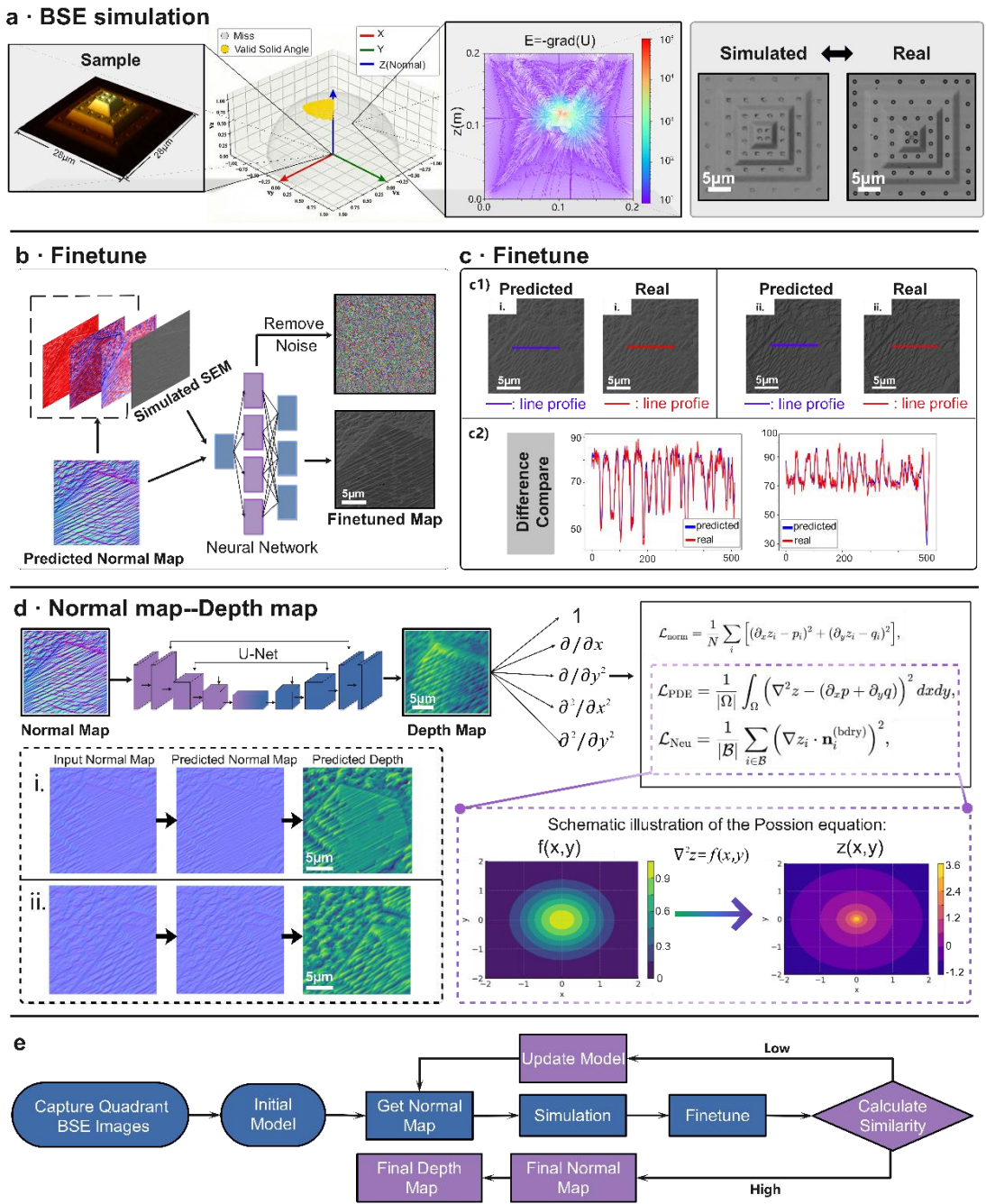

**Fig. 3 | Theoretical framework of electron-optical simulation and physics-data coupled reconstruction.**

**(a)**, First-principles-inspired BSE simulation pipeline. A 3D sample topography is taken as input; the detector geometry determines the admissible/valid emission region in angle space, the electrostatic field (visualized as E=−∇U) governs electron trajectories, and the resulting simulated BSE image is compared with a real measurement (simulated ↔ real). **(b)**, Residual "finetune" module for sim-to-real calibration. The predicted normal map and the physics-rendered (simulated) SEM/BSE image are fed into a lightweight neural network; the network suppresses structured bias/noise ("Remove Noise") and outputs a finetuned appearance map. **(c)**, Finetune effectiveness. Predicted (after finetune) and real images are juxtaposed for two examples **(c1)**, and their agreement is quantified by line profiles **(c2)** under identical sampling lines (predicted vs real). **(d)**, Normal-to-depth conversion with physics constraints. A U-Net-based mapper produces a depth map from the normal map, while derivative operators and loss terms highlight how normal consistency, PDE residual, and boundary constraints are enforced; illustrative examples **(i–ii)** show input normal → refined normal → predicted depth, alongside a schematic Poisson-equation view. **(e)**, Closed-loop analysis-by-synthesis workflow. Quadrant BSE images initialize a model to obtain a normal map; forward simulation and finetune produce a rendered image that is compared to measurements to compute similarity, and the similarity signal drives model updates until convergence, yielding final normal and depth maps.

Building on the hardware reference and the AI acceleration strategy described above, this section explains the core physical mechanism that supports ECLIPS: the practical implementation of electron-optics simulation, and a data-driven calibration mechanism that corrects the bias introduced by physical approximations. **Fig. 3** presents the full chain from first-principles modeling, to residual-learning correction, and finally to depth-field integration.

We first build a numerical simulation framework based on electrostatics and electron-scattering physics (**Fig. 3a**). The process is divided into two consecutive stages: interaction–emission and trajectory tracing–detection. The simulation takes a 3D surface-topography matrix as input (**Fig. 3a1**). In the interaction–emission stage, we adopt a single-scattering approximation model[4, 49], and assume that under normal incidence of the primary beam, the emission angles of backscattered electrons (BSE) follow the Lambertian Cosine Law[50]. Given that, at the micro/nanoscale, local surface-normal variations typically remain within 60°, we further introduce a "local invariance of the diffuse component" assumption to reduce computational cost while preserving physical plausibility[23].

We then move to the trajectory tracing–detection stage. Based on the calibrated imaging geometry, we solve the Laplace equation $\nabla^2 U = 0$ to obtain the electric potential distribution $U(x, y, z)$ between the pole piece and the specimen, and then derive the electric field as $E = -\nabla U$, as visualized in **Fig. 3a2**. Within this field, we numerically integrate the equations of motion for electrons emitted at different angles and trace their trajectories. In the coordinate system shown in **Fig. 3a3**, the yellow region indicates the Valid Solid Angle—the range of initial emission angles that allows electrons to reach the detector's sensitive area. By scanning all pixels on the specimen surface and integrating the Lambertian distribution over the Valid Solid Angle, we generate a theoretical BSE contrast image. **Fig. 3a4** compares the simulated image (left) with the experimentally acquired image (right). The strong agreement in geometric structures and shadow distributions establishes a physical foundation for the downstream learning-based inversion.

Despite capturing the dominant imaging behavior, BSESM still introduces unavoidable systematic bias when handling complex 3D surfaces, due to the diffuse-component approximation. This bias appears as nonlinear distortion in local contrast, and is fundamentally linked to multiple-scattering effects and non-Lambertian behavior on complex topographies.

To correct this theoretical error without breaking the global constraints of the physical model, we model the bias as structured noise and propose a physics–data dual-driven residual calibration ("Finetune") mechanism (**Fig. 3b**). In this design, the BSESM simulation provides the global physical backbone, while a lightweight neural network takes the predicted normal map and the initial simulated image as inputs, and is trained end-to-end to implicitly learn and compensate the systematic bias.

The improvement is shown clearly in **Fig. 3c**. After Finetune, the corrected simulated image (**Fig. 3c1 Predicted**) closely matches the experimental data (**Fig. 3c1 Real**) in both intensity distribution and texture details. The line-profile comparison (**Fig. 3c2**) further confirms that the corrected curve (blue) aligns with the real signal (red), effectively closing the final sim-to-real gap.

After accurate reconstruction of the 2D normal field, the final challenge is to convert it into a metrologically meaningful 3D Depth Map (**Fig. 3d**). Mathematically, this corresponds to integrating the normal field. However, classical approaches (such as Poisson solvers or path integration) are extremely sensitive to noise at the micro/nanoscale, often causing low-frequency drift or artifacts[29].

To address this, we build a physics-informed neural network (PINN) integrator. Instead of relying on a purely mathematical solver, the network is supervised using spatial coordinate pairs generated from a large set of AFM data, and directly predicts the depth scalar field. Crucially, we inject physical constraints explicitly into the loss function, including: PDE Residual: enforcing zero curl of the depth gradient to satisfy integrability. Normal Consistency: ensuring the surface normals derived from the predicted depth align with the input normal field. Neumann Boundary Condition: regularizing derivative behavior at the boundary.

This design combines the precision of data supervision with the robustness of physics constraints. As shown in **Fig. 3d(i, ii)**, even for surfaces with very small height-to-width ratios, the normals computed back from the predicted depth (middle) match the input normals (left), and the final depth map (right) is free of artifacts and drift.

In summary, **Fig. 3e** outlines the overall ECLIPS workflow: starting from 4Q-BSE image acquisition, a U-Net produces an initial normal field; VFT/DSM performs physics-based simulation followed by Finetune correction; and a similarity loss between simulated and measured images is computed. This error signal drives iterative model updates until convergence. This self-supervised closed loop based on analysis-by-synthesis ensures that the final normal field and depth map not only fit the data, but also strictly follow the physical laws of electron optics[51].

**Achieving metrological accuracy and closed-loop physical consistency validation**

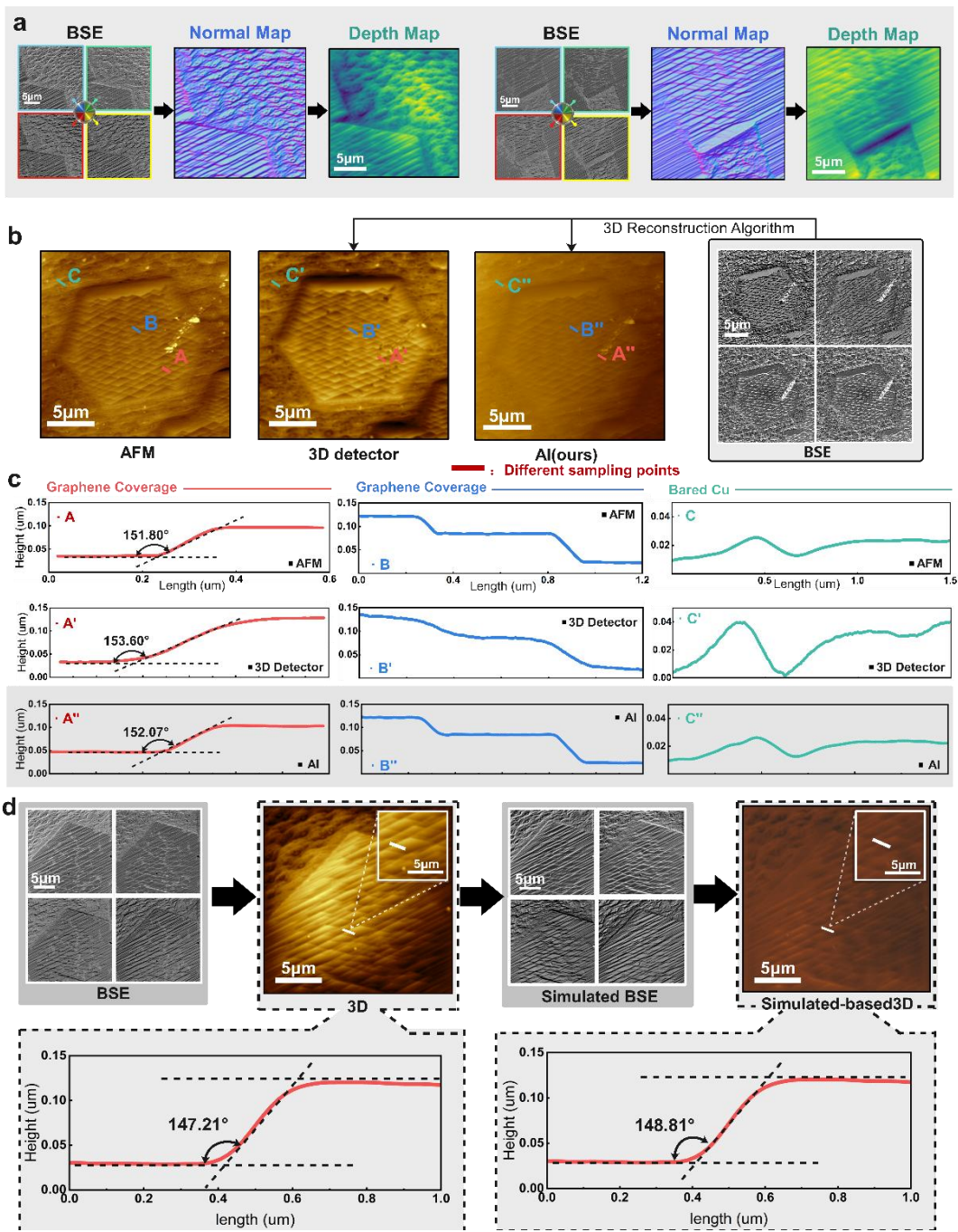

**Fig. 4 | Metrological validation on static nanostructures and closed-loop self-consistency.**

**(a)**, Generalization on unseen samples. For each test case, the 4Q-BSE input (2×2 mosaic) is first mapped to a normal map and then converted to a depth map (BSE → normal → depth), illustrating the end-to-end inference chain on structures not used for training. **(b)**, Cross-modal benchmarking on a graphene/Cu surface region. AFM provides the reference topography (left), a commercial 3D detector reconstruction is shown for comparison (middle), and the outputs corresponding 3D map of ECLIPS (right); colored labels (A/B/C and their counterparts) indicate consistent sampling locations across modalities, while the associated 4Q BSE images of the same field are shown on the far right. **(c)**, Profile-based metrology.

Height profiles extracted at the labeled positions compare AFM, commercial 3D detector, and ECLIPS(gray background), including representative step/terrace geometry (e.g., slope/step-angle annotations). **(d)**, Cycle-consistency test. Real 4Q BSE images are reconstructed into a 3D map, which is then forwarded through the simulator to synthesize BSE images; re-running the same reconstruction on the simulated BSE yields a second 3D map (real BSE → 3D → simulated BSE → simulated-based 3D), and the paired line profiles quantify the self-consistency.

Using the ECLIPS framework, we achieve end-to-end reconstruction from 4Q-BSE contrast images to 3D surface topography, and we validate its generality and robustness on multiple nanoscale samples with representative topological features. As shown in **Fig. 4a**, we apply the trained model to unseen samples that were not included during training: step-bunching structures on graphene-on-copper surfaces, as well as complex topographies with amorphous oxide textures. The figure presents two complete reconstruction examples, including the input 4Q-BSE contrast images, the predicted normal maps, and the derived depth maps. The results show that the framework preserves extremely high fidelity for both low-frequency undulations in flat regions and high-frequency discontinuities at step edges, demonstrating strong generalization across diverse nanoscale morphologies.

To assess metrology-level accuracy, we perform a quantitative multimodal comparison using the graphene–copper-foil sample as an example (**Fig. 4b–c**). We benchmark our reconstructed 3D topography against atomic force microscopy (AFM, used as ground truth) and a widely used commercial 3D detector solution. The rendered surfaces (**Fig. 4b**) show that the commercial 3D-detector software introduces noticeable smoothing and noise artifacts when handling micro/nanoscale details, leading to blurred step edges. In contrast, ECLIPS reconstructs sharp edge geometry, and its overall flatness is almost indistinguishable from the AFM height field. The cross-sectional height profiles in **Fig. 4c** further quantify this advantage: our reconstruction closely overlaps with AFM measurements, with the step-angle deviation controlled within $0.2°$, which is substantially better than the $1–2°$ systematic error of the commercial detector solution. These results confirm that physics-embedded deep learning not only mitigates the ill-posedness of the inverse problem, but also reaches metrology-grade performance in geometric quantification and edge preservation.

We further design a closed-loop reconstruct–simulate–reconstruct validation, a cycle-consistency test, to verify the self-consistency between ECLIPS and the physical imaging process (**Fig. 4d**). The procedure is as follows: we first reconstruct an initial depth map from real BSE images (**Depth Map 1**), feed it into the electron-optics forward model (BSESM) to generate simulated BSE images, and then send the simulated images back into the same reconstruction network to obtain a second depth map (**Depth Map 2**). Comparative analysis shows that the two depth maps—one from real signals and the other from simulated signals—are highly consistent at the pixel level. The recovered 3D structures differ only by negligible random perturbations, and their height-profile curves almost perfectly overlap. This closed-loop test demonstrates that our method establishes a robust bijective mapping between the 2D BSE imaging space and 3D surface topography, effectively avoiding the systematic biases commonly seen in traditional one-way modeling. It further confirms that the reconstruction is physically constrained, laying a solid foundation for subsequent in situ tracking of non-equilibrium dynamics.

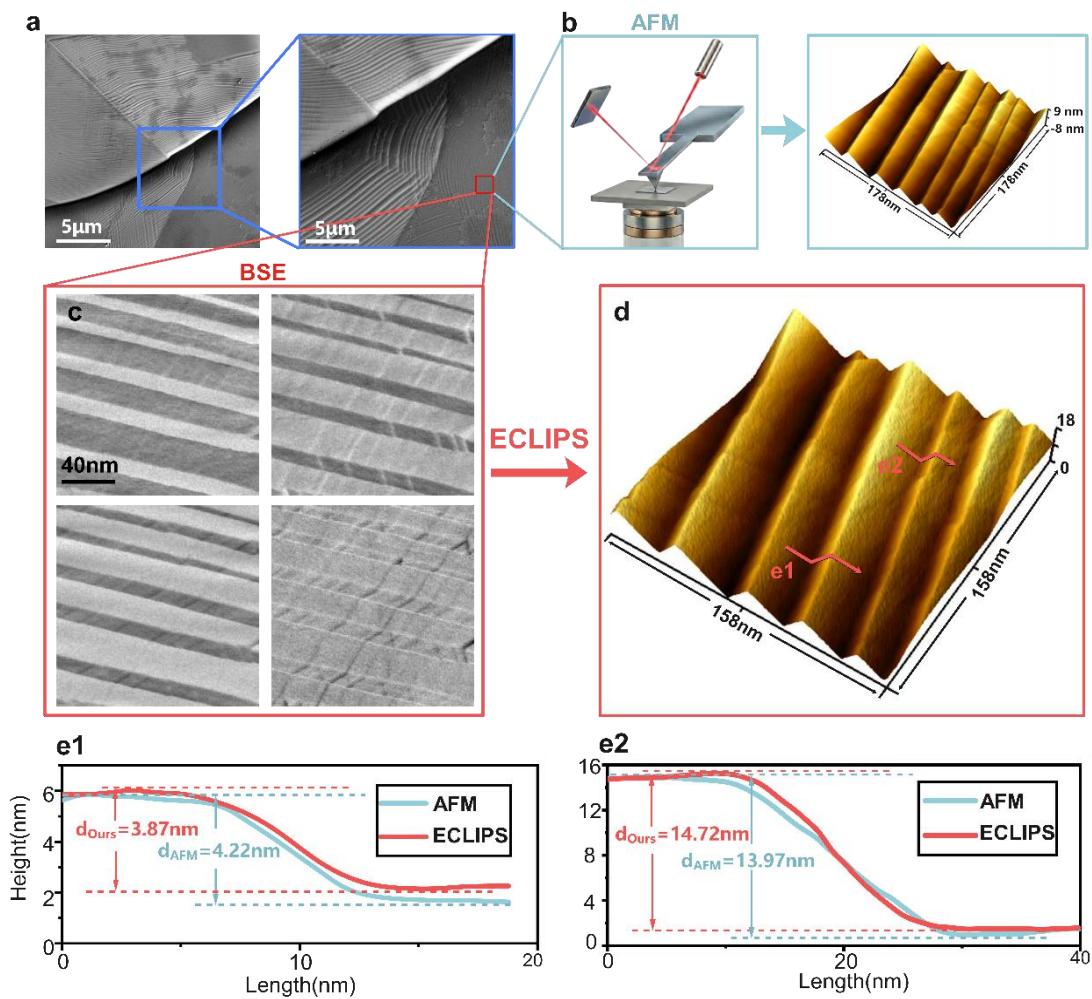

**Fig. 5 | Physics-embedded 3D SEM nanometrology at the step-edge limit.**

**(a)**, SEM overview and zoom-in localization. A wide-field SEM image is progressively magnified to the target region containing step-bunch features for quantitative evaluation (scale bars as indicated). **(b)**, AFM reference measurement of the corresponding region. The AFM setup schematic indicates the cross-modal acquisition, and the rendered AFM topography provides the height "ground truth" for comparison. **(c)**, Single-scan 4Q-BSE observations of the same region, providing multi-azimuth contrast cues of step edges and terraces (each quadrant image corresponds to one detector sector). **(d)**, Reconstructed 3D height map from the 4Q-BSE input using ECLIPS; the marked traces **(e1, e2)** indicate the locations where cross-sectional profiles are extracted. **e1–e2**, Line-profile comparisons along the two marked traces, showing agreement between the reconstructed height and the AFM reference across step transitions.

To rigorously define the capability limits of ECLIPS in accuracy, stability, and spatial resolution, we need a nanoscale "ruler" sample that (i) has a clear geometric ground truth, and (ii) appears naturally under SEM conditions and can be repeatedly measured. With this in mind, we choose graphene-induced step bunches on copper surfaces as our validation platform. These structures are widespread across large fields of view, and locally they exhibit regular layered step geometries. Their formation is also constrained by crystallography, giving them a rare combination of being both measurable and mechanistically interpretable—making them an ideal benchmark for testing whether a 3D inversion truly reaches metrology-grade reliability.

As shown in **Fig. 5a**, we first locate step-rich regions within a field of view on the order of 20 μm, and then progressively zoom in to a representative nanoscale field of view. This hierarchical zooming is not merely "framing the image"; it reflects a key requirement for practical microscopy: rapidly finding targets at the macro scale on complex surfaces, and then performing high-fidelity 3D metrology at the local scale—within the same closed-loop imaging and modeling system. We then scan the same region using contact-mode AFM to obtain a height matrix as a direct topographic reference (**Fig. 5b**). In parallel, the 4Q-BSE detector acquires four BSE contrast images with different directional information in a single scan (**Fig. 5c**). These complementary directional-shadow cues provide boundary conditions for electron-optics forward modeling and serve as the observation inputs to the reconstruction network.

Feeding the 4Q-BSE data into the ECLIPS framework yields the corresponding 3D height matrix (**Fig. 5d**). The reconstruction matches the AFM height field not only in step orientation, step density patterns, and overall undulation, but—more importantly—also shows clear geometric turning points and stable slope faces at step edges. This avoids a common failure mode of empirical mappings or weakly constrained inversions, where the global shape may look plausible but local edge geometry is overly smoothed. To push this agreement from "visual similarity" to metrological consistency, we extract and compare cross-sectional profiles from two representative regions (**Fig. 5e1, 5e2**). The results show that the reconstructed profiles reproduce the step-height changes and edge-transition features observed by AFM, demonstrating that the framework not only recovers low-frequency topography but also preserves sufficient sharpness and structural realism in high-frequency geometry.

Importantly, by comparing the spatial detail visible in the raw BSE contrast images with that in the reconstructed depth map, we confirm that ECLIPS does not introduce an additional scale-shrinking mechanism. Under our SEM settings, the effective imaging resolution is ~1.1 nm (see SI); correspondingly, the reconstructed depth map resolves geometric detail down to a comparable scale (~1.1 nm). This alignment indicates that the achievable spatial resolution is not dominated by algorithmic priors or numerical regularization, but is primarily limited by the intrinsic imaging resolution and sampling conditions of the SEM. In other words, under physically consistent constraints, ECLIPS avoids systematic bias against high-frequency geometry and prevents "algorithmic de-sharpening". The recoverable detail scale is determined by the effective information limit provided by the instrument itself, which mechanistically supports high-resolution, low-bias 3D metrology reconstruction at nanoscale step edges.

**Unveiling 4D spatiotemporal dynamics of surface structure**

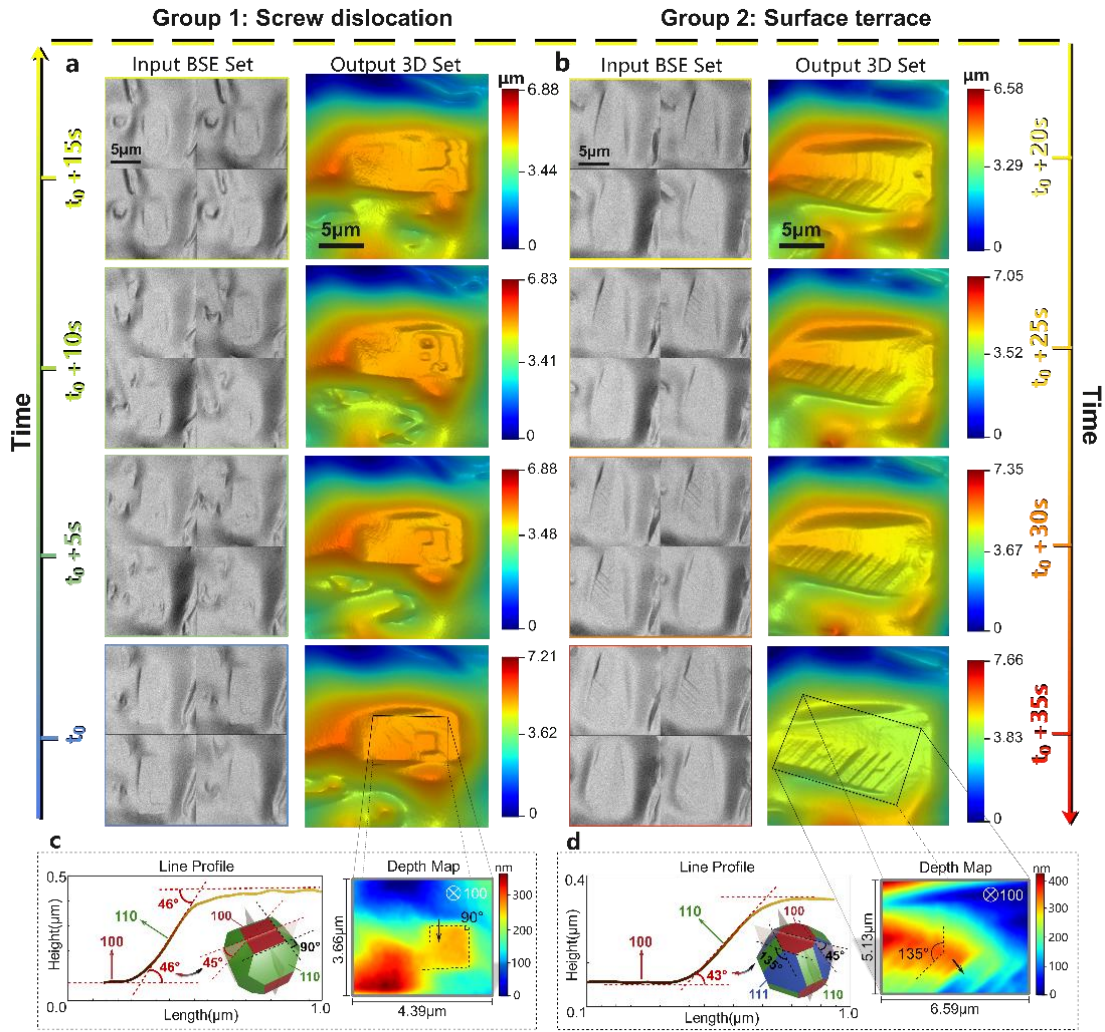

**Fig. 6 | In-situ 4D (3D + time) nanometrology of redox-driven Cu surface evolution.**
**(a)**, Group 1 (screw dislocation): time-resolved 4Q BSE inputs and reconstructed 3D outputs. For each time point (rows from $t_0$ to $t_0 + 15s$), the left panel shows the input BSE set (4Q mosaic), which is mapped by the reconstruction pipeline to the corresponding output 3D height map on the right (Input BSE Set → Output 3D Set), enabling frame-by-frame tracking of the defect-associated spiral topography. **(b)**, Group 2 (surface terrace): the same in-situ reconstruction applied to a terrace/step-bunch region over later frames (rows from $t_0 + 20s$ to $t_0 + 35s$); each row pairs the measured 4Q BSE mosaic with the reconstructed 3D map (Input → Output), showing consistent geometry over time. **(c)**, Crystallography-guided geometric readout for Group 1 at a selected frame/region (linked from the boxed area). A height line profile is extracted (arrow) to measure facet/step angles, and the inset Wulff polyhedron summarizes the low-index facet configuration in $Cu_2O$ consistent with the measured angles; the accompanying map visualizes in-plane directional relations (e.g., orthogonality) within the analyzed patch. **(d)**, Analogous analysis for Group 2 at a later time point (linked from the boxed area). A line profile quantifies terrace slopes/dihedral angles, while the in-plane intersection/angle map (e.g., ~135°) and Wulff-based facet annotation provide a crystallographic consistency check for the time-resolved 3D reconstruction.

After establishing geometric accuracy, physical consistency, and fine spatial resolution for static topography, we further extend the framework to probe non-equilibrium surface-evolution dynamics and assess its potential for in situ "4D (3D + time)" characterization. Crucially, BSE imaging can be performed under realistic gaseous environments, enabling direct observation of reaction-induced contrast variations during ongoing surface processes. This capability allows us to monitor solid–gas interfacial chemical reactions in their intrinsically non-equilibrium states and to quantify the accompanying 3D morphological evolution over time.

As shown in **Fig. 6**, we use the dynamic redox process on copper surfaces as a model system[32]. We perform continuous 3D monitoring of reaction-induced microstructural evolution under 121 Pa ($H_2 : O_2 = 5 : 1$). For each frame in the video sequence, we reconstruct 3D topography from the 4Q-BSE images using ECLIPS, and introduce spatial registration to obtain a dynamic 3D trajectory that is spatially self-consistent and temporally continuous[39, 52]. This turns conventional in situ electron-microscopy videos—traditionally limited to surface work-function-related contrast—into data with trackable, quantifiable 3D metrological meaning.

As shown in **Fig. 6a–b**, we reconstruct eight consecutively acquired sets of 4Q-BSE contrast images and analyze two representative crystallographic structures. **Group 1** focuses on $Cu_2O$ spiral dislocations that are extremely difficult to identify in BSE contrast. In conventional 2D BSE images, the dislocation-core region (**Fig. 6a Input BSE Set**) appears only as a very weak local contrast disturbance, easily buried by background noise and texture variations. With ECLIPS, these subtle signals are greatly amplified and presented as clear 3D structures: the time-resolved height maps reveal the characteristic spiral step morphology around the dislocation core, where steps rise in a helical pattern along the dislocation line. The topological location, spiral sense, and geometric relation to lattice orientation become immediately visible in 3D (**Fig. 6a Output 3D Set**), enabling a qualitative jump from "hidden in 2D" to "revealed in 3D". More importantly, the reconstruction does not merely detect the presence of steps; it stably tracks subtle frame-to-frame changes in local curvature, step height, and step spacing, providing a direct observational window into defect-driven surface dynamics.

**Group 2** shows the evolution of $Cu_2O$ steps formed by more visually apparent surface oxidation (**Fig. 6b Input BSE Set**). Unlike spiral-dislocation morphology, step bunches are often easier to notice in 2D contrast. However, the key for understanding surface-energy evolution is not qualitative recognition, but whether one can stably quantify step migration, slope angles, and local rearrangements. The time series in **Fig. 6** shows that ECLIPS can continuously reconstruct the 3D morphology of the step bunches during the surface reaction (**Fig. 6b Output 3D Set**), while maintaining consistent geometric scale and directionality across time. This advances step evolution from qualitative description to measurable, time-dependent facet parameters.

To further verify the physical realism and accuracy of ECLIPS 4D reconstruction, we perform cross-checks in **Fig. 6c–d** using geometric cross-sectional analysis for the two structure types (**Fig. 6c** at time t0, and **Fig. 6d** at time t0 + 35 s). We use the low-index facet combinations and dihedral-angle relations implied by the $Cu_2O$ Wulff construction as crystallographic constraints for validation[53].

For the spiral-dislocation-related structure (**Fig. 6c**), the reconstructed depth map shows an approximately square plateau near the dislocation core. The in-plane boundaries are mutually orthogonal, indicating a clear fourfold symmetry. To connect this in-plane geometry to real facet orientation, we select a line cut perpendicular to one plateau edge (black arrow in **Fig. 6c**) and extract a height line profile. We then fit a local tangent over the plateau-to-sidewall transition to obtain an effective slope angle. The characteristic angle from this profile (about 134° when expressed as an external angle, corresponding to a 46° inter-slope angle) is consistent with the dihedral-angle relation between the {100} top facet and the {110} side facet in the $Cu_2O$ Wulff construction. Together with the geometric constraints of a square top facet (the in-plane symmetry of {100}) and equivalent surrounding sidewalls, this morphology can be attributed to a crystallographic configuration with a {100} top surface bounded circumferentially by {110} side facets. In this way, the reconstructed dislocation-related steps and slopes are not only geometrically self-consistent, but also satisfy the facet orientations and angle relations expected from crystallography-selective growth of $Cu_2O$.

For the surface-step structure (**Fig. 6d**), the reconstructed depth map shows two step families intersecting in-plane with an intersection angle of about 135°. We extract a height line profile within the selected region along one step direction (black arrow in **Fig. 6d**), and fit a local tangent on the slope segment to obtain an effective slope angle. This slope angle follows the same geometric relation implied by the ~135° external-angle feature. Jointly comparing the in-plane intersection angle and the cross-sectional slope angle with the $Cu_2O$ Wulff construction suggests that this crossed-step geometry can be explained by a {100} top facet together with side facets jointly constrained by the {110} and {111} low-index facet families—namely, a {100}-up configuration where {110}/{111} facets participate in forming the corresponding ridges and intersection angles. Therefore, both the measured step intersection angle and the profile-derived slope angle in **Fig. 6d** fall within the facet-combination and dihedral-angle relations allowed by the Wulff construction. This crystallography-based analysis further supports the physical trustworthiness and accuracy of the time-series reconstructions produced by ECLIPS.

In summary, **Fig. 6** demonstrates frame-by-frame 3D reconstruction with ECLIPS during an in situ redox process. On one hand, the reconstructed sequence tracks the temporal evolution of two topological structure types—spiral dislocations and steps/terraces—within a unified coordinate system. On the other hand, the line-profile geometry is consistent with facet and dihedral-angle constraints from the $Cu_2O$ Wulff construction, providing crystallography-based physical validation and supporting the use of these reconstructions for extracting kinetic parameters and enabling mechanistic analysis.

**Conclusion and outlook**

In this work, we have bridged the historical divide between rigorous physical modeling and data-driven inference to establish a unified framework for quantitative nanometrology. By embedding the fundamental laws of electron optics directly into the gradient flow of deep neural networks, we convert the ill-posed inverse problem of 3D SEM reconstruction into a deterministic and tractable optimization task, enabling metrological, nanoscale 3D topography to be recovered from a single acquisition. Quantitatively, the achieved reconstruction resolution reaches 1.1 nm, indicating that the method is nearing the hardware-limited regime under the SEM settings.

The development of the VFT and the integration of PINN proved that physical priors—when formulated as differentiable operators—can effectively govern the behavior of "black box" models, ensuring that computational efficiency does not come at the cost of scientific confidence.

The implications of this framework extend far beyond static topographic mapping. The realization of 4D *in situ* tracking offers a transformative tool for materials science, allowing researchers to quantitatively observe and measure non-equilibrium dynamics—such as defect migration, phase transitions, and surface reactions—in real-time and real-space. This capability to visualize "the invisible" dynamics of defects like screw dislocations provides the missing link between microscopic mechanisms and macroscopic material performance.

Motivated by this technical strategy, ECLIPS suggests a broad horizon of opportunities. First, the framework is naturally extensible to other scattering-based imaging modalities, such as transmission electron microscopy (TEM), X-ray tomography, or optical scattering, where complex wave-matter interactions similarly hinder quantitative inversion. Second, future iterations could incorporate quantum mechanical scattering potentials into the forward model, potentially pushing reconstruction precision toward the atomic scale.